\documentclass[twocolumn,showpacs,preprintnumbers,amsmath,amssymb]{revtex4}

\usepackage{graphicx}
\usepackage{dcolumn}
\usepackage{bm}
\usepackage{soul}
\usepackage{color}
\usepackage{epstopdf}
\usepackage[version=3]{mhchem}
\begin{document}

\title{Surface plasmon resonances of protein-conjugated gold nanoparticles \\ on graphitic substrates}

\author{ Anh D. Phan$^{1,2}$, Trinh X. Hoang$^{2}$, Thi H. L. Nghiem$^{2}$, Lilia M. Woods$^{1}$}
\affiliation{$^{1}$Department of Physics, University of South Florida, Tampa, Florida 33620, USA}%
\email{anhphan@mail.usf.edu}
\affiliation{$^{2}$Institute of Physics, 10 Daotan, Badinh, Hanoi, Vietnam}%

\date{\today}

\begin{abstract}
We present theoretical calculations for the absorption properties of protein-coated gold nanoparticles on graphene and graphite substrates. As the substrate is far away from nanoparticles, numerical results show that the number of protein bovine serum molecules molecules aggregating on gold surfaces can be quantitatively determined for gold nanoparticles with arbitrary size by means of the Mie theory and the absorption spectra. The presence of graphitic substrate near protein-conjugated gold nanoparticles substantially enhances the red shift of the surface plasmon resonances of the nanoparticles. Our findings show that graphene and graphite provide the same absorption band when the distance between the nanoparticles and the substrate is large. However at shorter distances, the resonant wavelength peak of graphene-particle system differs from that of graphite-particle system. Furthermore, the influence of the chemical potential of graphene on the optical spectra is also investigated.

\end{abstract}

\maketitle

At nanoscale, the enhancement of surface interactions and quantum confinement leads to significant property difference between bulk materials and nanoparticles (NPs) and provides peculiar applications. Magnetic NPs made of $\ce{Fe_3O_4}$ or $\ce{Co}$ are strongly sensitive to external fields. While noble metal NPs, particularly gold NPs (AuNPs), are gaining considerable attention due to their fascinating optical properties derived from their localized surface plasmon resonances. NPs, therefore, have been widely used in a variety of devices such as solar cells \cite{1,2,3}, electrocatalysts \cite{4}, and sensors with high sensitivity \cite{5,6,7}. In many applications, the support of substrates to metallic NPs are proven to be highly effective in maximizing the performance of these structures \cite{8,9,10,11,12}. One of the most interesting features induced by the presence of a substrate is the red shift in the resonance wavelength. This effect exhibits more strongly in metallic NPs than in other types of NPs and are very useful for designing optical sensors. 
\begin{figure}[htp]
\includegraphics[width=8.3cm]{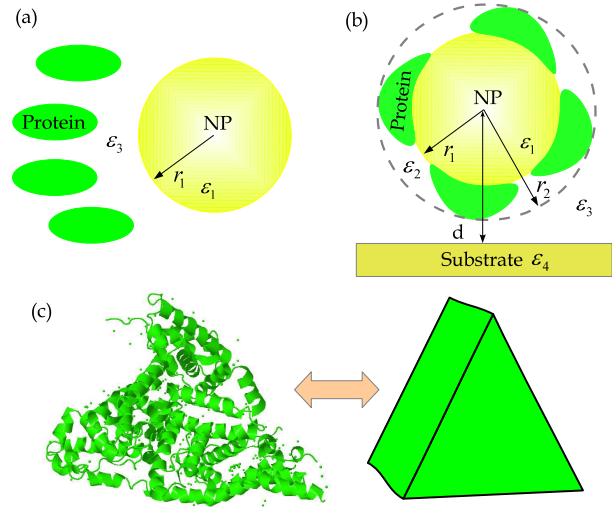}
\caption{\label{fig:1}(Color online) (a) BSA protein adsorbed in the AuNP. (b) Schematic illustration of AuNP and substrate with the separation distance $d$ counted from the center of NP to the substrate surface. (c) BSA protein can be modelled by an equilateral triangular prism with dimensions 84 $\times$ 84 $\times$ 84 $\times$ 31.5 $\AA$.}
\end{figure}

Graphene, a one-atom thick material in nature, with outstanding mechanical, optical and electric properties has been broadly investigated both theoretically and experimentally as a promising support material for metallic NPs \cite{8,9,10,13}. In theoretical studies, researchers used the dielectric function of pristine graphene fitted from previous experiments \cite{32}. However, this kind of dielectric function cannot be used to study the change of the optical response of graphene due to chemical doping or external fields. In this paper we present theoretical calculations for the absorption and scattering cross section of AuNPs in an aqueous protein BSA solution with a graphene substrate illustrated in Fig.\ref{fig:1}a. The graphene dielectric function can be calculated using the Kubo formulism which allows us to investigate the influence of chemical potential of graphene on the optical properties of AuNPs. Calculations for the substrate made of graphite are studied to compare with the graphene substrate interactions. Bovine serum albumin (BSA) protein plays a crucial role in suppressing the assembly of AuNPs due to the van der Waals interactions after these NPs are always synthesized in a $\ce{HAuCl_4}$ solution \cite{33}. BSA proteins adhering on the AuNP surface reduce the particle-particle interactions. Another biological function of BSA is the ability to bind numerous substances in the body. One can apply this feature for delivering and attaching AuNPs to biological sites in living bodies \cite{33}.

The Mie theory has been widely employed to calculate the scattering and absorption components of the extinction spectrum of an object. The expression of the scattering and absorption cross section are given \cite{14}
\begin{eqnarray}
\sigma_{sca}=\frac{k^4}{6\pi}|\alpha |^2, \nonumber\\
\sigma_{abs}=kIm(\alpha),
\label{1}
\end{eqnarray}
where $\alpha$ is the polarizability of the object, $k=2\pi n/\lambda$ and $n$ is the refractive index of medium. To obtain overview about the polarizability of the isolated object, we consider the core-shell particle structure described in Fig. \ref{fig:1}b without the presence of substrate (d = $\infty$). For this geometry, the polarizability $\alpha$ is written by \cite{14}
\begin{eqnarray}
\alpha = 4\pi r_2^3\frac{\varepsilon_2\varepsilon_a-\varepsilon_3\varepsilon_b}{\varepsilon_2\varepsilon_a+2\varepsilon_3\varepsilon_b},\\
\varepsilon_a=\varepsilon_1\left[1+2\left(\frac{r_1}{r_2}\right)^3\right]+2\varepsilon_2\left[1-\left(\frac{r_1}{r_2}\right)^3\right],\\
\varepsilon_b=\varepsilon_1\left[1-\left(\frac{r_1}{r_2}\right)^3\right]+\varepsilon_2\left[2+ \left(\frac{r_1}{r_2}\right)^3\right],
\label{2}
\end{eqnarray}
where $\varepsilon_1$, $\varepsilon_2$, and $\varepsilon_3$ are the dielectric function of core (AuNP), shell and surrounding medium, respectively. These expressions were derived from the Maxwell-Garnet (MG) theory for an effective medium approximation. The dielectric function of gold is always modelled by Drude model $\varepsilon_1 = 1-\omega_p^2/[\omega(\omega+i\Gamma)]$, here $\omega_p$ is the plasma frequency of gold and $\Gamma$ is the damping parameter. This model provides a close approximation for bulk gold materials in some cases. Such a model, however, predicts the surface plasmon resonance of AuNP in vacuum to be around 216 $nm$ while this resonant wavelength is observed around 510 $nm$ \cite{15}. The calculation fails to describe the optical properties of AuNP due to an underestimation of the size effect and the role of bound electrons in NPs. In this paper, we used the Lorentz-Drude model for the gold dielectric response \cite{16}
\begin{eqnarray}
\varepsilon_1(\omega)=1-\frac{f_0\omega_p^2}{\omega(\omega +i\Gamma_0)}+\sum_{j}\frac{f_j\omega_p^2}{\omega_j^2-i\omega\Gamma_j-\omega^2},
\label{3}
\end{eqnarray}
in which $f_0$ and $f_j$ are the oscillator strength corresponding to frequency $\omega_j$ and the damping parameter $\Gamma_j$. The fist two terms in Eq.(\ref{3}) represent for the Drude model in the dielectric function of gold. Other terms involve the interband transition due to bound electron contribution modelled by the Lorentz oscillator. To be convenient for investigating the absorption versus wavelength, the dielectric function $\varepsilon_1(\omega)$ can be rewritten in terms of wavelength with new parameters: the plasma wavelength - $\lambda_p$, the damping wavelength - $\gamma_j$, and the interband transition wavelength - $\lambda_j$. The values of these parameters are given in Table \ref{table:1}.
\begin{eqnarray}
\varepsilon_1(\lambda)&=&1-\frac{f_0/\lambda_p^2}{1/\lambda^2 +i/(\lambda\gamma_0)}\nonumber\\
&+&\sum_{j=1}^5\frac{f_j/\lambda_p^2}{1/\lambda_j^2 -i/(\lambda\gamma_j)-1/\lambda^2}.
\label{4}
\end{eqnarray}
\begin{table}[htp]
\caption{Parameters for dielectric function of AuNP and graphite provided in Ref.\cite{16,17}}
\centering 
\begin{tabular}{c c c} 
\\[0.5ex]
\hline\hline 
Parameter & AuNP & Graphite \\ [0.5ex] 
\hline 
$f_0$ & 0.76 & 0.014 \\ 
$\gamma_0$ (nm) & 23438.9 & 195.17 \\
$\lambda_p$ (nm)& 138 & 46.01 \\
$f_1$ & 0.024 & 0.073 \\ 
$\gamma_1$ (nm) & 5154.6 & 302.84 \\
$\lambda_1$ (nm)& 2993.4 & 4517.31 \\
$f_2$ & 0.010 & 0.056 \\ 
$\gamma_2$ (nm) & 3600.75 & 169.52 \\
$\lambda_2$ (nm)& 1496.7 & 354.122 \\  
$f_3$ & 0.071 & 0.069 \\ 
$\gamma_3$ (nm) & 418.41 & 878.54 \\
$\lambda_3$ (nm)& 1427.9 & 279.10 \\  
$f_4$ & 0.601 & 0.005 \\ 
$\gamma_4$ (nm) & 498.1 & 27005.65 \\
$\lambda_4$ (nm)& 288.63 & 91.403 \\  
$f_5$ & 4.384 & 0.262 \\ 
$\gamma_5$ (nm) & 561.1 & 667.164 \\
$\lambda_5$ (nm)& 93.26 & 87.323 \\  
$f_6$ & --- & 0.460 \\ 
$\gamma_6$ (nm) & --- & 104.2 \\
$\lambda_6$ (nm)& --- & 15.55 \\  
$f_7$ & --- & 0.2 \\ 
$\gamma_7$ (nm) & --- & 31.78 \\
$\lambda_7$ (nm)& --- & 38.81 \\  
[0.5ex] 
\hline 
\end{tabular}
\label{table:1} 
\end{table}

The dielectric function of gold in Eq.(\ref{4}) was measured for metallic film \cite{16}. However, to describe more accurately the size effect of nanoparticle, particularly when the particle's diameter is less than 20 $nm$, it is necessary to change the scattering frequency $\Gamma_0$ to $\Gamma = \Gamma_0+Av_F/r_1$, here $v_F$ is the Fermi velocity of gold \cite{18}, A is the parameter including the scattering processes with the magnitude ranging from 0.1 to 1 \cite{18}. It is easy to see that the parameter A presents the influence of the size effect on the AuNP dielectric function.

After obtaining AuNPs in water, we have to mix the AuNP solution with aqueous solution of bovine serum albumin (BSA) described in Fig.\ref{fig:1}a. BSA proteins and AuNPs attract each other and the proteins finally stick to the gold surface. The protein layer leads to the shell with dielectric function $\varepsilon_2$ shown in Fig.\ref{fig:1}b. For this geometry, we can determine $\varepsilon_2$ as a composite medium consisting of protein and water
\begin{eqnarray}
\varepsilon_2(\lambda)=f\varepsilon_{protein}+(1-f)\varepsilon_w,
\label{5}
\end{eqnarray}
where $f$ is the percentage of protein in the shell and $\varepsilon_w = 1.77$ is the dielectric constant of water. Note that it is assumed that there is no change in the dielectric in the visible spectrum. $\varepsilon_{protein}$ is the dielectric function of protein \cite{19} as a function of wavelength
\begin{eqnarray}
\varepsilon_{protein}(\lambda)=1+\sum_{j}\frac{1/\Lambda_j^2}{1/\lambda_j^2-i/(\lambda\gamma_j)-1/\lambda^2},
\label{6}
\end{eqnarray}
in which $\Lambda_1=10853.54$ $nm$, $\Lambda_2=878.5$ $nm$, $\Lambda_3=92.6$ $nm$, $\Lambda_4=82.81$ $nm$, $\gamma_1=\infty$, $\gamma_2=2484.52$ $nm$, $\gamma_3=155.28$ $nm$, $\gamma_4=65.38$ $nm$, $\lambda_1=6059.8$ $nm$, $\lambda_2=194.1$ $nm$, $\lambda_3=99.38$ $nm$ and $\lambda_4=57.78$ $nm$.
\begin{figure}[htp]
\includegraphics[width=8.5cm]{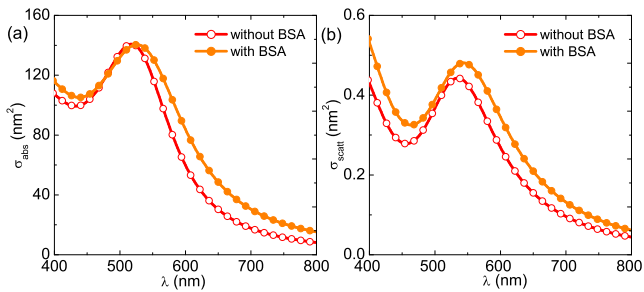}
\caption{\label{fig:2}(Color online) The absorption and scattering cross section of AuNPs in water with and without BSA in visible spectrum.}
\end{figure}

Figure \ref{fig:2} shows the absorption and scattering cross section of AuNPs submerged in water and BSA solvent. AuNPs are always designed to have the diameter 16 $nm$, and the thickness of nanoshell is experimentally estimated to be 3.15 $nm$ \cite{22}. It suggests that only a protein monolayer is formed on the NP surface. In the case of pure water medium, the parameter A is 0.4 to achieve the absorption curves matching with the curve in Ref. \cite{15}. Meanwhile A = 0.61 for the case of aqueous protein solutions. The increase of A illustrates that the BSA-NP binding enhances the role of the size effect. The surface plasmon peak of NPs in solvent with and without BSA has a similar magnitude. As given in Eq.(\ref{1}), $\sigma_{abs}$ strongly depends on the imaginary part of $\alpha$ that is easily tuned by changing A. This figure also reveals that the absorption is dominant in comparison with the scattering. The extinction, thus, can be identified by the absorption. As a result, the absorption is the focus of examination in our next calculations. 

Apart from A, another important factor to determine the red shift in the plasmon resonant frequency is $f$ since it directly influences the dielectric function $\varepsilon_2$. This parameter is also used to calculate the amount of protein forming the nanoshell of AuNPs. Authors in Ref. \cite{21} indicated that the surface plasmon maximum shifts from 519 $nm$ \cite{26} to 526 $nm$ as proteins were added in aqueous solution. BSA proteins locating around NPs cause the increase of phase retardation through creating a medium of higher dielectric constant. To achieve these data sets, we take $f=0.4$. Another research \cite{27} obtained the same shift of surface plasmon absorption peak when mixing AuNPs with DNA solution. As a result, the impact of both DNA and BSA on the absorption properties of AuNPs are equal.

The average number $N$ of protein molecules binding on the surface of AuNPs can be calculated by
\begin{eqnarray}
N=\frac{4\pi f(r_2^3-r_1^3)}{3V_0},
\label{7}
\end{eqnarray}
where $V_0$ is the volume of a BSA protein molecule. In neutral solution, it has been shown that the conformation of BSA is well-described by a heart-shaped structure \cite{22,23,24} sketched in Fig.\ref{fig:1}c. Therefore $V_0$ = 96.24 $nm^3$ and N $\approx$ 15. Our value of N compares well with previous experiments \cite{22}. Authors in Ref.\cite{22} observed 12 protein molecules coating on AuNP that has a diameter 13 $nm$. 

In practical, it is difficult to synthesize all NPs with the same size. It was found that the size effects has a considerable impact on the optical response of NP \cite{26}. Individual metallic NPs with desired shape and size with attached BSA proteins can be characterized using optical trapping \cite{34,35}. From Eq.(\ref{7}) and experimental data from previous study \cite{22}, we have $f = 0.4$ for AuNP $r_1 = 6.9$ and 29.1 $nm$, respectively, while $f = 0.45$ when $r_1 = 15.2$ $nm$. It suggests the fraction $f$ is around the value of 0.4 and can be approximately constant as changing the radius. In addition, BSA molecules only assemble a layer on the AuNP surfaces. One can find that the increase of the NP size causes the increase of the number of protein molecules binding on the gold surface.
\begin{figure*}[htp]
\includegraphics[width=15.5cm]{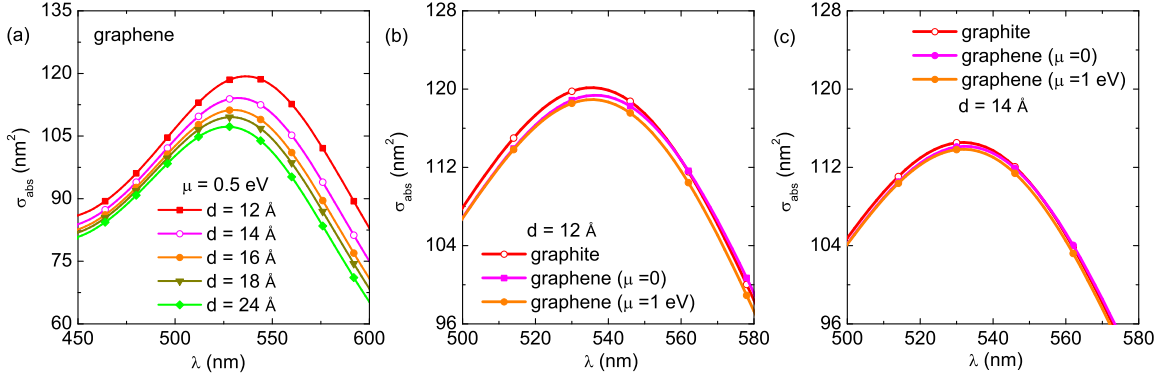}
\caption{\label{fig:3}(Color online) The absorption cross section of protein-coated AuNPs in an aqueous solution on (a) graphene substrate with $\mu = 0.5$ eV with different value of $d$, (b) graphite, pristine graphene and doping graphene with $\mu = 1$ eV at d = 12 $\AA$, and (c) graphite, pristine graphene and doping graphene with $\mu = 1$ eV at d = 14 $\AA$.}
\end{figure*}

As a substrate is brought close to AuNP (see Fig.\ref{fig:1}c), the polarizability of particles should be modified to be \cite{20}
\begin{eqnarray}
\alpha_m = \alpha\left[1-\left(\frac{r_2}{2d}\right)^3\frac{\varepsilon_2\varepsilon_a-\varepsilon_3\varepsilon_b}{\varepsilon_2\varepsilon_a+2\varepsilon_3\varepsilon_b}\frac{\varepsilon_4-\varepsilon_3}{\varepsilon_4+\varepsilon_3}  \right]^{-1},
\label{8}
\end{eqnarray}
where $\varepsilon_4$ is the dielectric function of the substrate. It is important to understand the dielectric function $\varepsilon_4$ to study the substrate effect on the polarizability of NPs. In our work, we consider the substrate made of graphene and graphite. The dielectric function of graphite is expressed by Eq.(\ref{4}) and parameters can be found in Table \ref{table:1}. These data sets are derived from \emph{ab initio} calculation \cite{16}. Meanwhile, the expression of the dielectric function of graphene is given by \cite{25}
\begin{eqnarray}
\varepsilon_g(\omega) &=& 1+\frac{i\sigma(\omega)}{\varepsilon_0\omega h_0},\label{9}\\
\sigma(\omega)&=&\frac{2ie^2k_BT}{\pi\hbar^2\omega}\ln\left(2\cosh\frac{\mu}{2k_BT}\right)\nonumber\\
&+&\frac{e^2}{4\hbar}\left[\theta(\hbar\omega-2\mu)-\frac{i}{2\pi}\ln\frac{(\hbar\omega+2\mu)^2}{(\hbar\omega-2\mu)^2}\right],
\label{10}
\end{eqnarray}
in which $h_0=0.34$ $nm$ is the thickness of graphene, $\varepsilon_0$ is the vacuum permittivity, $k_B$ is the Boltzmann constant, $T$ is temperature, $\hbar$ is the Plank constant, $e$ is the charge of electron, $\sigma(\omega)$ is the graphene optical conductivity which can be described by the Kubo formalism, $\theta$ is the step function and $\mu$ is the chemical potential of graphene. The factor $1/h_0$ represents the number of layers within the distance unit. One may use Eq.(\ref{9}) to characterize the dielectric function for graphite because both Eq.(\ref{4}) and Eq.(\ref{9}) take into account the effect of multiple layers on the optical properties. Note that Eq.(\ref{10}) is valid in the case of doping graphene with $\mu \gg k_BT$. For pristine graphene ($\mu = 0$), we can take $\sigma(\omega)=\sigma_0=e^2/4\hbar$.

Evidence from Fig.\ref{fig:3}a suggests that graphene with $\mu = 0.5$ $eV$ has a substantial effect on the absorption of BSA-conjugated AuNPs at short distances. The surface plasmon resonance frequency is 536.4 and 532.3 $nm$ at d = 12 and 14 $\AA$, respectively. The impact becomes weaker as the separation distance $d$ increases. For d = 24 $\AA$, a peak of the absorption spectrum appears at 527.1 $nm$. It suggests that when d $\geq$ 25 $\AA$, graphene is almost isolated from AuNPs.

Figure \ref{fig:3}b and \ref{fig:3}c illustrate the influence of chemical potential of graphene as well as the effect caused by graphene and another carbon structure, namely graphite, on the optical spectrum of protein-coated AuNPs. It can be seen that for d = 12 $\AA$, the absorption curve of graphene with $\mu = 1$ $eV$ only deviates from pure graphene when $\lambda  \geq 528$ $nm$. The maximum absorption for graphene with $\mu = 1$ $eV$ and $\mu = 0$ takes place at 535.6 $nm$ and 536.5 $nm$, respectively. One, therefore, can conclude that the increase of chemical potential of graphene gives rise to the blue shift in the surface plasmon resonance frequency. The modification, however, is minor and it is hard to detect in experimentation. While as the substrate is graphite, the absorption peak is located near 535.5 $nm$. We can observe the change in the amplitude of $\sigma_{abs}$ in comparison with the case of graphene substrate. Interestingly, the atomic structure of graphene and graphite plays an important role in the optical spectra at d = 12 $\AA$ and the separation between curves disappears at d = 14 $\AA$ (Fig.\ref{fig:3}c). There is no discrepancy between graphene and graphite at longer range.

The chemical potential of graphene can be tailored by chemical doping or applying external fields. The relation between an applied electric field and chemical potential is expressed by \cite{28,29}
\begin{eqnarray}
\frac{\pi\varepsilon_0\hbar^2v_0^2}{e}E=\int_0^{\infty}E\left[f(E)-f(E+2\mu)\right]dE,
\label{11}
\end{eqnarray}
where $f(E)$ is the Fermi distribution function, $v_0 =c/300$ is the graphene Fermi velocity and $c$ is the speed of light. From Eq.(\ref{11}), one finds that the pristine graphene can obtain $\mu = 1$ $eV$ if an electric field $E = 6.64\times 10^9$ $V/m$ is applied to graphene sheet. According to recent research \cite{30,31}, there is no major change in structure and properties of protein induced by both weak and normal electric field. The electric fields may not profoundly affect the absorption cross section of protein-conjugated AuNPs.

In conclusion, the optical properties of AuNPs in the protein solution have been theoretically studied using the Mie theory. Adsorption of BSA to AuNPs weakens the van der Waals interactions between NPs and generates the stabilization of AuNPs. Our calculations explains the red shift in the plasmon resonance due to BSA. Moreover, it is important to note that directly observing the number of protein molecules binding to NPs is still a remarkable challenge. However, this number can be calculated precisely using the absorption spectrum and the model of the effective dielectric function. We also study the impact of graphene and graphite substrate on $\sigma_{abs}$. The substrate effect results in the shift in the surface plasmon resonance wavelength. The similar curves were observed in the absorption spectra of BSA-coated AuNPs in the case of graphene and graphite substrate at $d \geq 14$ $\AA$, but the big difference between two curves arises at shorter distances $d$. Since chemical potential of graphene does not change the absorption while $\mu$ strongly depends on the electric field and the chemical doping, $\sigma_{abs}$ may not be influenced by both this external field and doping.
\begin{acknowledgments}
We gratefully acknowledge helpful discussions with Prof. Onofrio M. Marago. This work was supported by the Nafosted Grant No. 103.01-2013.16. Lilia M. Woods acknowledges the Department of Energy under contract DE-FG02-06ER46297.
\end{acknowledgments}

\end{document}